\documentclass[onecolumn,english,prd]{revtex4}
\usepackage{graphicx}
\begin{document}

\newcommand{\lexp}{\mathop{\langle}}
\newcommand{\rexp}{\mathop{\rangle}}
\newcommand{\rexpc}{\mathop{\rangle_c}}
\newcommand{\beq}{\begin{equation}}
\newcommand{\eeq}{\end{equation}}
\newcommand{\beqa}{\begin{eqnarray}}
\newcommand{\eeqa}{\end{eqnarray}}

\def\k{{\hbox{\bf k}}}
\def\q{{\hbox{\bf q}}}
\def\x{{\hbox{\bf x}}}
\def\r{{\hbox{\bf r}}}
\def\dD{\delta_{\rm D}}

\def\Mpc{\, h^{-1} \, {\rm Mpc}}
\def\Gpc{\, h^{-1} \, {\rm Gpc}}
\def\Gpccube{\, h^{-3} \, {\rm Gpc}^3}
\def\kvecMpc{\, h \, {\rm Mpc}^{-1}}
\def\la{\mathrel{\mathpalette\fun <}}
\def\ga{\mathrel{\mathpalette\fun >}}
\def\fun#1#2{\lower3.6pt\vbox{\baselineskip0pt\lineskip.9pt
        \ialign{$\mathsurround=0pt#1\hfill##\hfil$\crcr#2\crcr\sim\crcr}}}
% \title{A New Approach to Gravitational Clustering}
% \title{The Growth of Structure in Renormalized Perturbation Theory}
% \title{Renormalized Perturbation Theory and Gravitational Clustering}
\title{Renormalized Cosmological Perturbation Theory}
\author{Mart\'{\i}n  Crocce and  Rom\'an Scoccimarro}
\vskip 2pc
\address{Center for Cosmology and Particle Physics,\\
Department of Physics, New York University, New York, NY 10003}

\begin{abstract}

We develop a new formalism to study nonlinear evolution in the growth of large-scale structure, by following the dynamics of gravitational clustering as it builds up in time. This approach is conveniently  represented by Feynman diagrams constructed in terms of three objects: the initial conditions (e.g. perturbation spectrum), the vertex (describing non-linearities) and the propagator (describing linear evolution). We show that loop corrections to the linear power spectrum organize themselves into two classes of diagrams: one corresponding to mode-coupling effects, the other to a renormalization of the propagator. Resummation of the latter gives rise to a quantity that measures the memory of perturbations to initial conditions as a function of scale. As a result of this, we show that a well-defined (renormalized) perturbation theory  follows, in the sense that each term in the remaining mode-coupling series dominates at some characteristic scale and is subdominant otherwise.  This is unlike standard perturbation theory, where different loop corrections can become of the same magnitude in the nonlinear regime.  In companion papers we compare the resummation of the propagator with numerical simulations, and apply these results to the calculation of the nonlinear power spectrum. Remarkably, the expressions in renormalized perturbation theory can be written in a way that closely resembles the halo model.

\end{abstract}

\maketitle

\section{Introduction: improving perturbation theory}

The growth of density and velocity perturbations is a fundamental result of cosmology that can be used to test basic properties of the universe, such as the amount of dark matter and dark energy. Although the linear and weakly non-linear growth of perturbations is well understood,  giving the large-scale asymptotics (tree-level amplitudes in diagrammatic perturbation theory language~\cite{treePT}) of the power spectrum and higher-order statistics as a function of time, the nonlinear corrections to these results (``loop corrections"~\cite{loopPT}) are less well understood (see~\cite{PTreview} for a review).

The standard approach of perturbation theory (hereafter PT) is to solve the equations of motion by expanding in powers of the perturbation amplitudes. This is well justified when doing tree-level PT, as one is essentially looking for asymptotic behavior (the large-scale limit, where perturbation amplitudes become very small compared to unity), however as one approaches the nonlinear scale and loop corrections become increasingly important, the validity of such a method becomes less clear. Part of the issue is that cosmological PT is not  a standard PT in the sense of having a small coupling constant; its validity depends on the scale, redshift, type of initial perturbation spectrum and statistic under consideration. For example, for initial spectra with significant large-scale power (see Fig.~13 in~\cite{PTreview}) or at high redshift for CDM models, including just one-loop corrections to the linear power spectrum gives very good agreement with numerical simulations. On the other hand, for redshifts $z \approx 0$ when the effective spectral index at the nonlinear scale becomes $n_{\rm eff} \simeq -1$ one-loop PT deviates up to $20\%$ for $k \leq 0.2~\kvecMpc$ (see Fig.~6 in~\cite{zdist}). Similar results hold for the bispectrum~\cite{SCFFHM98}.

The purpose of this paper is to cast PT in a different form, which helps to understand the physical meaning of different contributions in the infinite perturbation series, following ideas initially put forward in~\cite{Sco00}. In particular we show that different class of diagrams organize themselves into a few characteristic quantities, the most important of which (in terms of improving PT) is the non linear propagator that measures the memory of perturbations to their initial conditions as a function of scale. When an infinite number of diagrams is resummed into this quantity, the remaining (infinite number of) diagrams form a well-defined expansion where each term with $\ell$ loops corresponds to the effects of $\ell+1$-mode-coupling, and dominates in a narrow range of scales.

\begin{figure*}
\begin{center}
\begin{tabular}{cc}
{\includegraphics[width=0.5\textwidth]{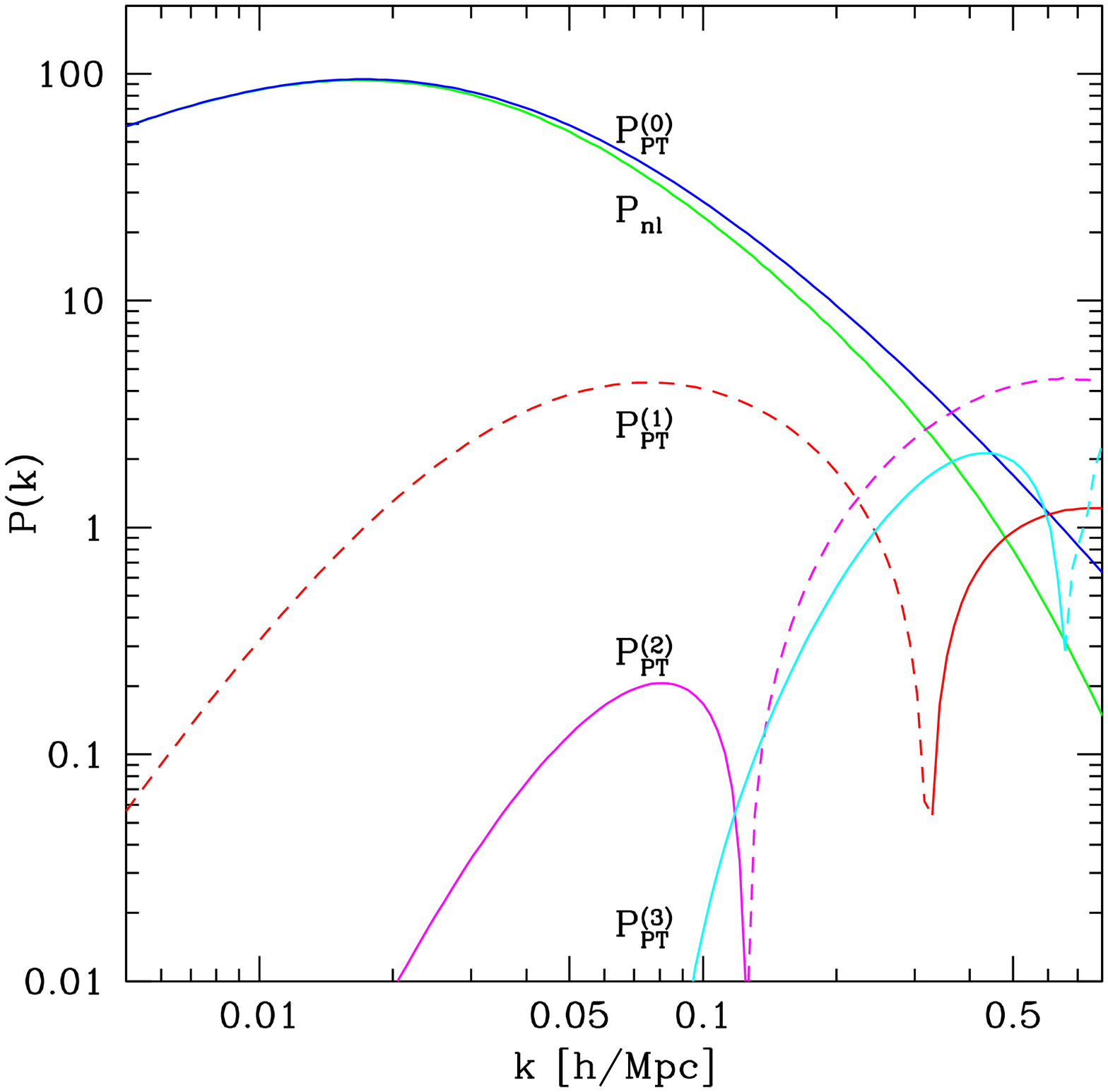}}&
{\includegraphics[width=0.5\textwidth]{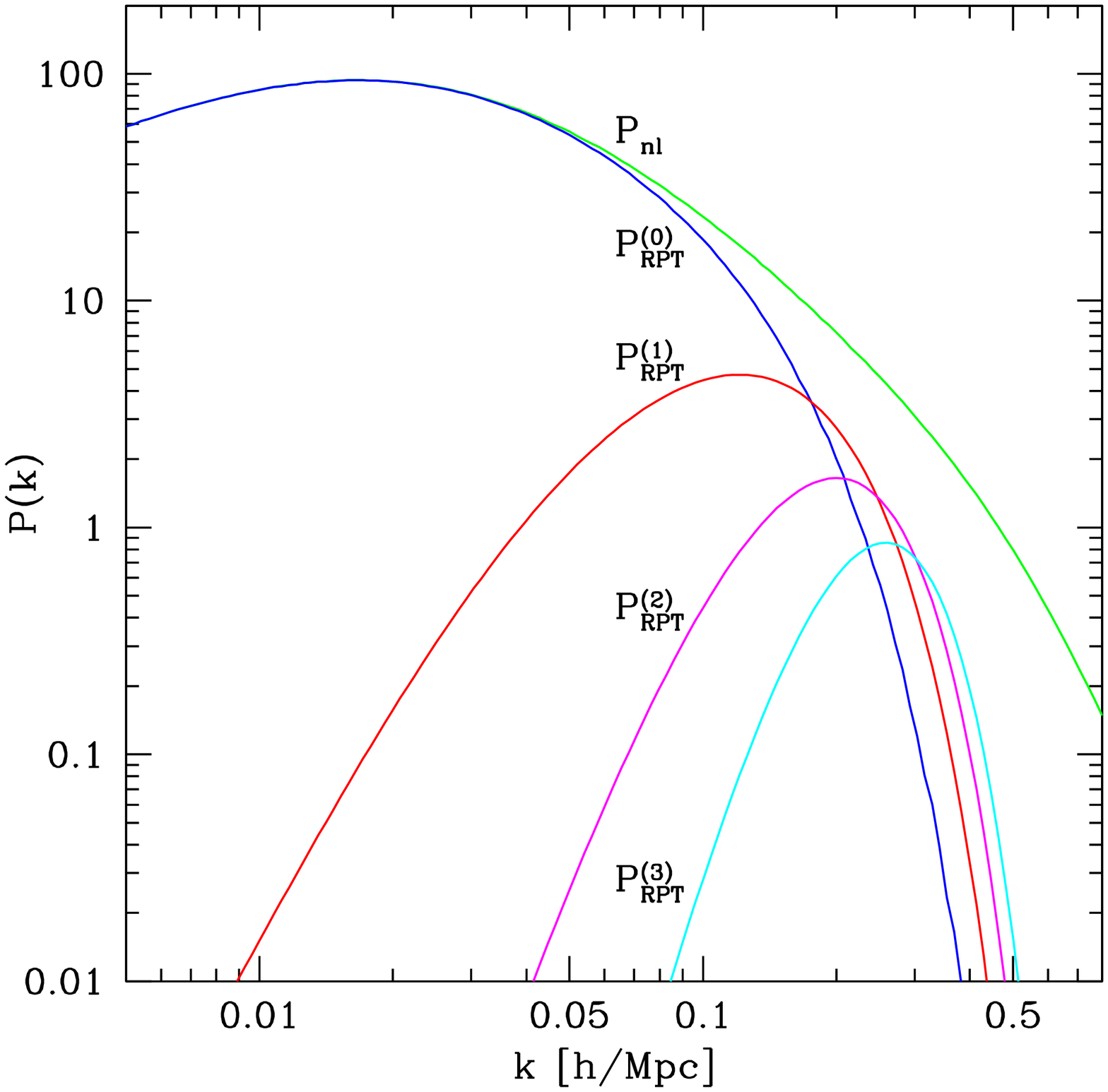}}
\end{tabular}
\caption{Comparison between PT (left) and RPT (right) loop expansion for the nonlinear power spectrum in the Zel'dovich approximation.  $P_{\rm nl}$ denotes the exact result for the nonlinear power spectrum, Eq.~(\protect\ref{PZA}), whereas $P_{\rm PT}^{(\ell)}$ ($P_{\rm RPT}^{(\ell)}$) denotes the $\ell$-loop correction in PT (RPT). Dashed lines denote negative values. The resummation of the propagator involved in RPT leads to a well-defined perturbation expansion even in the nonlinear regime, unlike PT.}
\label{PTvsRPT}
\end{center}
\end{figure*}

To illustrate these ideas and hopefully motivate the reader to go through the rest of the paper (which is rather   technical for those not familiar with field theory methods), a quick calculation of the nonlinear power spectrum in the Zel'dovich approximation provides a good understanding of the difference in behavior at nonlinear scales between (standard)  PT and the renormalized perturbation theory (RPT hereafter) we develop in this paper. In the Zel'dovich approximation~\cite{ZA}, the trajectories of particles away from their initial (Lagrangian) positions $\q$ are given by $\x=\q+\Psi(\q)$ where  $\Psi(\q)$ is the displacement field. The matter density perturbation, defined as $\delta(\x)\equiv\rho(\x)/\bar\rho-1$, can be written in Fourier space as,
 
\beq
\delta(\k)\equiv\int \frac{d^3x}{(2\pi)^3}\ \delta(\x)\ {\rm e}^{i\k \cdot \x}=\int \frac{d^3q}{(2\pi)^3}\ {\rm e}^{i\k \cdot \q}\ \Big[ {\rm e}^{i\k \cdot \Psi(\q)}-1\Big],
\label{prima}
\eeq
by means of the mass conservation relation $[1+\delta(\x)] d^3x=d^3q$. This gives for the power spectrum, defined as $P(k)\ \delta_{\rm D}(\k+\k')\equiv \lexp \delta(\k)\delta(\k') \rexp $,

\beq
P(k)=\int \frac{d^3r}{(2\pi)^3}\ {\rm e}^{i\k \cdot \r}\ \Big[\langle {\rm e}^{i\k \cdot \Delta\Psi}\rangle-1\Big],
\label{seconda}
\eeq
where we introduced $\r \equiv \q-\q'$, $\Delta\Psi \equiv \Psi(\q)-\Psi(\q')$ and used that $\langle {\rm e}^{i\k \cdot \Psi}\rangle=1$ from Eq.~(\ref{prima}). In the Zel'dovich approximation the displacement satisfies $\Psi=-i({\k}/k^2)\delta_L$, where $\delta_L$ stands for the linear density perturbation in Fourier space. Moreover, for Gaussian initial conditions $\Psi$  is a Gaussian random field and thus $\langle {\rm e}^{i\k \cdot \Delta\Psi}\rangle= {\rm e}^{-\frac{1}{2}\langle(\k \cdot \Delta\Psi)^2\rangle}$, which can be used to cast Eq.~(\ref{seconda}) as,

\beq
P(k)=\int \frac{d^3r}{(2\pi)^3}\ {\rm e}^{i\k \cdot \r}\ \Big[ {\rm e}^{-[k^2 \sigma_v^2-I(\k,\r)]}-1\Big],
\label{PZA}
\eeq
where $I(\k,\r)\equiv \int  d^3q\, (\k\cdot\q)^2\, \cos(\q\cdot\r) P_L(q)/q^4$ and $\sigma_v^2 = I(k,0)/k^2$ is the variance of the displacement field (and also the one-dimensional velocity dispersion in linear theory). Equation~(\ref{PZA}) is the well-known relationship between the nonlinear and linear power spectrum in the Zel'dovich approximation~\cite{PZA}. Although this provides a poor description of the power spectrum (with the nonlinear power being smaller than linear), having an exact solution such as Eq.~(\ref{PZA}) serves to illustrate the differences between PT and RPT. In PT, one basically expands in the amplitude of the linear spectrum, therefore

\beq
P(k)=\int \frac{d^3r}{(2\pi)^3}\ {\rm e}^{i\k \cdot \r}\ \sum_{n=1}^\infty \frac{(-1)^n}{n!} [k^2 \sigma_v^2-I(\k,\r)]^n \equiv \sum_{\ell=0}^\infty\ P_{\rm PT}^{(\ell)}(k),
\label{Ppt}
\eeq
where the term with $n=1$ gives the linear spectrum (tree-level, zero loops, i.e. $\ell=0$), $n=2$ gives one-loop ($\ell=1$) corrections, and so on. The left panel in Fig.~\ref{PTvsRPT} shows the $P_{\rm PT}^{(\ell)}(k)$ for $\ell=0,1,2,3$,  for a CDM model with normalization $\sigma_8=0.9$ and shape parameter $\Gamma=0.175$, with nonlinear scale $k_{\rm nl}\simeq 0.2 \kvecMpc $. Solid lines denote positive contributions, dashed lines negative ones. Note that at large enough scales only the linear spectrum ($\ell=0$) contributes, as expected, however as the nonlinear scale is approached different loop corrections become of the same order, with significant cancellations among them. Similar behavior has been seen in the case of PT for the exact dynamics, at least for some initial spectra, see e.g.~\cite{Fry94}.

In RPT, as we discuss in detail below, one attempts (among other things) to sum an infinite class of diagrams for the propagator, which (as will be shown in section~\ref{GZA}) correspond to the factor of ${\rm e}^{-k^2 \sigma_v^2}$ in the Zel'dovich approximation, Eq.~(\ref{PZA}). The remaining infinite series corresponds to the result

\beq
P(k)=\int \frac{d^3r}{(2\pi)^3}\ {\rm e}^{i\k \cdot \r}\ {\rm e}^{-k^2 \sigma_v^2}
 \sum_{n=1}^\infty \frac{[I(\k,\r)]^n }{n!} \equiv \sum_{\ell=0}^\infty\ P_{\rm RPT}^{(\ell)}(k),
\label{Prpt}
\eeq

which is shown in the right panel in Fig.~\ref{PTvsRPT} for $\ell=0,1,2,3$. We see that as a result of this partial resummation, the behavior of the resulting perturbation expansion is drastically altered: successive terms dominate at increasingly smaller scales, and there are no cancellations among them (all contributions are positive). Moreover, as we discuss below, RPT provides a physical understanding of the different contributions: the sum represents the contribution of $n$-modes coupling (the larger $n$ the generation of power moves to smaller scales), the decaying exponential factor represents how the amplitude and phases of the Fourier modes differ from linear evolution from the initial conditions. Indeed, it is easy to see that Eq.~(\ref{Prpt}) can be rewritten as

\beq
P(k)= {\rm e}^{-k^2 \sigma_v^2} \
 \sum_{n=1}^\infty n! \int \delta_{\rm D}(\k- \q_{1\dots n})\, \Big[ F_n(\q_1,\ldots,\q_n)\Big]^2
P_L(q_1)d^3q_1  \ldots P_L(q_n)d^3q_n,
\label{PrptKernels}
\eeq
where  $\q_{1\dots n}=\q_1+\ldots+\q_n$, and $F_n$ represents the PT kernels [see Eqs.~(\ref{ec:deltan}-\ref{Gn}) for a definition] in the Zel'dovich approximation [see Eq.~(\ref{FZA}) below]. The sum corresponds to all mode-coupling diagrams (with the appropriate $n!$ weighting), with all ``tadpole" diagrams in the language of standard PT~\cite{ScFr96} being summed to the exponential prefactor. This propagator renormalization factor is given by the square of $\langle{\cal D}\delta(\k)/{\cal D}\delta_{L}(\k)\rangle/ D_+= \exp(-k^2\sigma_v^2/2)$, where ${\cal D}$ represents here a functional derivative~\footnote{In the sections below we use $\delta$ to denote functional derivatives when there is no possibility of confusion with the density field perturbations.} and $D_+$ is the linear growth factor. The exponential decay is caused by the deviations from linear evolution (see section~\ref{GZA} for a derivation), and the decay length (here $\sigma_v^{-1}\simeq 0.16 \kvecMpc$) defines a characteristic scale at which individual Fourier modes do not evolve as in linear PT. The negative sign contributions in PT (see left panel in Fig.~\ref{PTvsRPT}) are an artifact of the power series expansion of propagator renormalization; as $k$ increases these terms become large and unless one sums up the infinite series this leads to a breakdown of PT. Note that the $n=1$ term in Eq.~(\ref{PrptKernels}) gives a contribution to $P(k)$ which is proportional to the linear spectrum, i.e. $ {\rm e}^{-k^2 \sigma_v^2} P_L(k)$, therefore if the linear spectrum were cutoff at some scale $k_c$ [$P_L(k)=0$ for $k>k_c$] the nonlinear effects encoded in propagator renormalization do not lead to power generation at $k>k_c$. It is in this sense that $n\geq 2$ encodes mode-mode coupling.

In summary,  the identification in RPT of physical quantities such as the propagator in the perturbation series helps to cure the lack of convergence of PT in the nonlinear regime. Notice also that the behavior of RPT in Fig.~\ref{PTvsRPT} is similar to the halo model description of the power spectrum, the similarities are explored below in section~\ref{halos}.

The rest of this paper is organized as follows. In the next section we review the standard results from PT and describe an alternative description necessary to develop RPT. Section~\ref{stats} applies the result in the previous section to the calculation of statistics, in particular the power spectrum and the nonlinear propagator. Section~\ref{RPT} derives the main results of RPT, and section~\ref{halos} discusses the connection between  RPT and the halo model. The conclusions are presented in section~\ref{conclude}.

\section{Dynamics}

\subsection{In Standard Form}

For completeness, we review here the standard form of the equations of motion in Fourier space and their solutions in standard PT, before  going into the new approach that helps us understand better the different terms in the perturbation series and thus facilitate the process of resummation.

We study the evolution of Cold Dark Matter (CDM) in the single-stream approximation, where the relevant equations of motion are conservation of mass, momentum, and the Poisson equation (see e.g.~\cite{Peebles80}). Consistent with this, the velocity field is assumed to be irrotational, then gravitational instability can be fully described in terms of the density contrast $\delta (\x,\tau)=\rho(\x,\tau)/\bar \rho - 1$ and the peculiar velocity divergence $\theta\equiv\nabla\cdot{\bf v}$. Defining the conformal time $\tau=\int dt/a$, with $a(\tau)$ the cosmological scale factor, and the conformal expansion rate ${\cal H}\equiv {d\ln a /{d\tau}}$, the equations of motion in Fourier space become~\cite{Fry84,GGRW86,JaBe94}

\beq
\frac{\partial \tilde{\delta}(\k ,\tau)}{\partial \tau}+\tilde{\theta}(\k,\tau) = - \int d^3 k_1 d^3 k_2 \, \delta_{\rm D}(\k-\k_1-\k_2)\,\alpha(\k_1, \k_2) \, \tilde{\theta}(\k_1,\tau)\, \tilde{\delta}(\k_2,\tau)   \label{ddtdelta},
\eeq
\beq
\frac{\partial \tilde{\theta}(\k,\tau)}{\partial \tau}+ {\cal H}(\tau)\tilde{\theta}(\k,\tau) + {3\over 2}\Omega_m (\tau) {\cal H}^2(\tau)\tilde{\delta}(\k,\tau) = - \int d^3k_1 d^3k_2\, \delta_{\rm D}(\k-\k_1-\k_2) \beta(\k_1, \k_2)\,\tilde{\theta}(\k_1,\tau)\,\tilde{\theta}(\k_2,\tau) \label{ddttheta}.
\eeq
The non linear interactions (in the right-hand side) are responsible for the coupling between different Fourier modes, given by

\beqa
\alpha(\k_1, \k_2) \equiv {(\k_1+\k_2) \cdot \k_1 \over{k_1^2}}\,,
\qquad \beta(\k_1, \k_2) \equiv {|\k_1+\k_2|^2 (\k_1 \cdot \k_2 )\over{2 k_1^2
k_2^2}}.
\nonumber
\eeqa
Equations (\ref{ddtdelta}) and (\ref{ddttheta}) are valid in an arbitrary homogeneous and isotropic universe, which evolves according to the Friedmann equations. For simplicity here we will restrict ourselves to an Einstein-de Sitter model for which $\Omega_m=1$ and $\Omega_{\Lambda}=0$.
As a consequence, these equations can formally be solved with the following perturbative
expansion,
\begin{equation}
\tilde{\delta}(\k,\tau) = \sum_{n=1}^{\infty} a^n(\tau)
\delta_n(\k),\ \ \ \ \ \tilde{\theta}(\k,\tau) = -
{\cal H}(\tau) \sum_{n=1}^{\infty} a^n(\tau) \theta_n(\k)
\label{ptansatz},
\end{equation}
where {\em only the fastest growing mode is taken into account} (below we discuss how to generalize this to the full time dependence).   The equations of motion, Eqs.~(\ref{ddtdelta}-\ref{ddttheta})
determine $\delta_n(\k) $ and $\theta_n(\k)$ in terms of the linear
fluctuations,
\begin{equation}
\delta_n(\k) = \int d^3\q_1 \ldots \int d^3\q_n\, \dD(\k -
\q_{1\ldots n}) F_n(\q_1, \ldots ,\q_n) \delta_0(\q_1) \ldots
\delta_0(\q_n) \label{ec:deltan},
\end{equation}
\begin{equation}
\theta_n(\k) = \int d^3\q_1 \ldots \int d^3\q_n\, \dD(\k -
\q_{1\ldots n}) G_n(\q_1, \ldots ,\q_n) \delta_0(\q_1) \ldots
\delta_0(\q_n) \label{ec:thetan},
\end{equation}
where $F_n$ and $G_n$ are homogeneous functions
of the wave vectors \{$\q_1, \ldots ,\q_n $\} with degree
zero. They are constructed from the fundamental mode coupling
functions according to the recursion relations~\cite{GGRW86}

\begin{eqnarray}
F_n(\q_1, \ldots ,\q_n) &=& \sum_{m=1}^{n-1} { G_m(\q_1, \ldots ,\q_m)
\over{(2n+3)(n-1)}} \Bigl[(2n+1) \alpha(\k_1,\k_2) F_{n-m}(\q_{m+1},
\ldots ,\q_n) \nonumber \\ & & +2 \beta(\k_1, \k_2)
G_{n-m}(\q_{m+1}, \ldots ,\q_n) \Bigr] \label{Fn},
\end{eqnarray}
\begin{eqnarray}
G_n(\q_1, \ldots ,\q_n) &=& \sum_{m=1}^{n-1} { G_m(\q_1, \ldots ,\q_m)
\over{(2n+3)(n-1)}} \Bigl[3 \alpha(\k_1,\k_2) F_{n-m}(\q_{m+1}, \ldots
,\q_n) \nonumber \\ & & +2n \beta(\k_1, \k_2) G_{n-m}(\q_{m+1},
\ldots ,\q_n) \Bigr] \label{Gn},
\end{eqnarray}
where $\k_1 \equiv \q_1 + \ldots + \q_m$, $\k_2 \equiv \q_{m+1} +
\ldots + \q_n$, $\k \equiv \k_1 +\k_2$, and $F_1= G_1 \equiv 1$, with $\delta_1(\k)=\theta_1(\k)=\delta_0(\k)$, the initial perturbations.

\subsection{In Matrix Form}

The equations of motion can be rewritten in a more symmetric form~\cite{Sco98,Sco00} by defining a two-component ``vector''

\beq
\Psi_a(\k,\eta) \equiv \Big( \delta(\k,\eta),\ -\theta(\k,\eta)/{\cal H} \Big),
\eeq
where the index $a=1,2$; and  we have introduced a new time variable,

\beq
\eta\equiv \ln a(\tau).
\eeq
corresponding to the number of e-folds of the scale factor. In the cosmology we are considering ($\Omega_m=1$), $a(\tau)$ represents also the linear growth factor of perturbations in the growing mode. Using $\Psi_a(\k,z)$ one can rewrite Eqs.~(\ref{ddtdelta}) and (\ref{ddttheta}) together as (we henceforth use the convention that repeated Fourier arguments are integrated over),

\beq
\partial_\eta \Psi_a(\k,\eta) + \Omega_{ab} \Psi_b(\k,\eta) =
 \gamma_{abc}^{(\rm s)}(\k,\k_1,\k_2) \ \Psi_b(\k_1,\eta) \ \Psi_c(\k_2,\eta),
\label{eom}
\eeq
where
\beq
\Omega_{ab} \equiv \Bigg[
\begin{array}{cc}
0 & -1 \\ -3/2 & 1/2
\end{array}        \Bigg],
\eeq
and $\gamma_{abc}^{(\rm s)}$ is the symmetrized {\em vertex} matrix given by

\beqa
\gamma_{121}^{({\rm s})}(\k,\k_1,\k_2)&=&\delta_{\rm D}(\k-\k_1-\k_2) \ \alpha(\k_1,\k_2)/2,
\label{ga121} \nonumber \\
\gamma_{112}^{({\rm s})}(\k,\k_1,\k_2)&=&\delta_{\rm D}(\k-\k_1-\k_2) \ \alpha(\k_2,\k_1)/2,
\label{ga112} \nonumber \\
\gamma_{222}^{({\rm s})}(\k,\k_1,\k_2)&=&\delta_{\rm D}(\k-\k_1-\k_2) \ \beta(\k_1,\k_2),
\label{ga222}
\label{vertexdefinition}
\eeqa
and zero otherwise, $\gamma_{abc}^{(\rm s)}(\k,\k_i,\k_j)=\gamma_{acb}^{(\rm s)}(\k,\k_j,\k_i)$. An implicit integral solution to Eq. (\ref{eom}) can be found by Laplace transforming in the variable $\eta$,

\beq
\sigma_{ab}^{-1}(\omega) \ \Psi_b(\k,\omega) = \phi_a(\k) +
\gamma_{abc}^{(\rm s)}(\k,\k_1,\k_2) \oint \frac{d\omega_1}{2\pi i} \
\Psi_b(\k_1,\omega_1) \Psi_c(\k_2,\omega-\omega_1),
\label{eom2}
\eeq
where  $\phi_a(\k)\equiv\Psi_a(\k,\eta=0)$ denotes the initial conditions, set when the growth factor is one,  and $\sigma_{ab}^{-1}(\omega) \equiv \omega \delta_{ab}+\Omega_{ab}$. Multiplying by the matrix,

\[
\sigma_{ab}(\omega) = \frac{1}{(2\omega+3)(\omega-1)} \Bigg[ \begin{array}{cc}
2\omega+1 & 2 \\ 3 & 2\omega \end{array} \Bigg].
\]
and performing the inversion of the Laplace transform, we finally get the formal solution

\beq
\Psi_a(\k,\eta) = g_{ab}(\eta) \ \phi_b(\k) + \int_0^\eta  d\eta' \ g_{ab}(\eta-\eta')
\ \gamma_{bcd}^{(\rm s)}(\k,\k_1,\k_2)\ \Psi_c(\k_1,\eta') \Psi_d(\k_2,\eta'),
\label{eomi}
\eeq
where the {\em linear propagator} $g_{ab}(\eta)$ is defined as ($c>1$ to pick out the standard retarded propagator~\cite{Sco98})

\beq
g_{ab}(\eta) = \oint_{c-i\infty}^{c+i\infty} \frac{d\omega}{2\pi i}
\sigma_{ab}(\omega) \ {\rm e}^{\omega \eta} = \frac{{\rm e}^\eta}{5}
\Bigg[ \begin{array}{rr} 3 & 2 \\ 3 & 2 \end{array} \Bigg] -
\frac{{\rm e}^{-3\eta/2}}{5}
\Bigg[ \begin{array}{rr} -2 & 2 \\ 3 & -3 \end{array} \Bigg],
\label{prop}
\eeq
for $\eta\geq 0$, whereas $g_{ab}(\eta) =0$ for $\eta<0$ due to causality, and $g_{ab}(\eta) \rightarrow \delta_{ab}$ as $\eta\rightarrow 0^{+}$. The {\em linear propagator} is the Green's function of the linearized version of Eq.~(\ref{eom}) and describes the standard linear evolution of density and velocity fields from {\em any} configuration of initial perturbations. From a physical point of view, the most interesting initial conditions are given by when $\delta(\k,\eta=0)$ and $\theta(\k,\eta=0)$ are proportional random fields (instead of independent), in which case we can write 

\beq
\phi_a(\k)=u_a \, \delta_0(\k)
\eeq
where $u_a$ is a two component ``vector''. This covers the usual case of {\em growing mode} initial conditions for which $u=(1,1)$, for which the second term in Eq.~(\ref{prop}) does not contribute upon contraction of $g_{ab}$ with $\phi_b(\k)$, and {\em decaying mode} initial conditions for which $u=(2/3,-1)$. For the sake of simplicity and definiteness, we will only contemplate the case $\phi_a(\k)=u_a \, \delta_0(\k)$ throughout this paper.

\subsection{Diagrammatic Representation of the Solution}
\label{DiagrammaticRepresentation}

Equation~(\ref{eomi}) for a given $\k$-mode has a simple interpretation. The first term corresponds to the linear propagation from the initial conditions, whereas the second term contains information on non-linear interactions (mode-mode coupling). This non-linear contribution comes from the interaction of all pairs of waves $\k_1$ and $\k_2$ (whose sum is $\k$ due to translational invariance), at all intermediate times $\eta'$ (with $0 \leq \eta' \leq \eta$). The interaction is characterized by the vertex $\gamma_{bcd}$ and then linearly evolved in time by $g_{ab}(\eta-\eta')$. As we will see shortly, this simple interpretation will lead, when applied recursively, to a graphical representation of the solution $\Psi_a$ in terms of diagrams that allows the resummation of the different statistical objects of interest.

With the help of Eq.~(\ref{eomi}), we seek for an explicit expression for $\Psi_a(\k,\eta)$ in the form of a series expansion [see Eq.~(\ref{ptansatz})]

\beqa
\Psi_a(\k,\eta)=\sum_{n=1}^{\infty} \Psi^{(n)}_a(\k,\eta),
\label{seriesexp}
\eeqa
with  [compare with Eqs.~(\ref{ec:deltan}-\ref{ec:thetan})]

\beq
\Psi_a^{(n)}(\k,\eta)= \int \delta_{\rm D}(\k-\k_{1\ldots n})
{\cal F}_a^{(n)}(\k_1,\ldots,\k_n;\eta) \delta_0(\k_1) \ldots \delta_0(\k_n),
\label{seriesol}
\eeq
where $\k_{1\ldots n} \equiv \k_1+ \ldots + \k_n$. Replacing Eqs.~(\ref{seriesexp},\ref{seriesol}) into Eq.~(\ref{eomi}), we find the recursion relation satisfied by the kernels  [see Eqs.~(\ref{Fn}-\ref{Gn})]

\beqa
& {\cal F}_a^{(n)}(\k_1,\ldots,\k_n;\eta) \ \delta_{\rm D}(\k-\k_{1\ldots n}) =  \nonumber \\ 
= & \left[\sum_{m=1}^n \int_0^\eta ds \,g_{ab}(\eta-s) \, \gamma_{bcd}^{(\rm s)}(\k,\k_{1 \ldots m},\k_{m+1 \ldots n}) \, {\cal F}_c^{(m)}(\k_{1 \ldots m};s) \, {\cal F}_d^{(n-m)}(\k_{m+1\ldots n};s)\right]_{\rm symmetrized},
\label{recursionkernels}
\eeqa
where the r.h.s. has to be symmetrized under interchange of any two wave vectors. For $n=1$, ${\cal F}_a^{(1)}(\eta)=g_{ab}(\eta)u_b$. Equation~(\ref{recursionkernels}) reduces to the standard recursion relations given by Eqs.~(\ref{Fn}-\ref{Gn}) in the limit that initial conditions are imposed in the infinite past, i.e. by replacing the lower limit of integration in Eq.~(\ref{recursionkernels}) by $s=-\infty$, in which case only the fastest growing mode survives and ${\cal F}^{(n)}=a^n\, (F_n,G_n)$. Otherwise Eq.~(\ref{recursionkernels}) gives the full time dependence of PT solutions, including all transients from initial conditions~\cite{Sco98}.

Although in principle the solution for the kernels gives us an analytic expression for $\Psi^{(n)}$, it quickly becomes very cumbersome with increasing $n$, therefore in practice this is not an ideal way to proceed; this is however the route taken in standard PT. Another reason apart from technical complication is the fact that the kernels (thought as functions of $n$) contain redundant information, since they all share the same building blocks, the vertex and linear propagator. In order to overcome these disadvantages we will introduce a graphical representation of the basic objects of the theory, and simple rules to combine them, such as to build a set of diagrams that are in one to one correspondence with each order $\Psi^{(n)}$.
 
\begin{figure}[h!]
\includegraphics[width=0.9\textwidth]{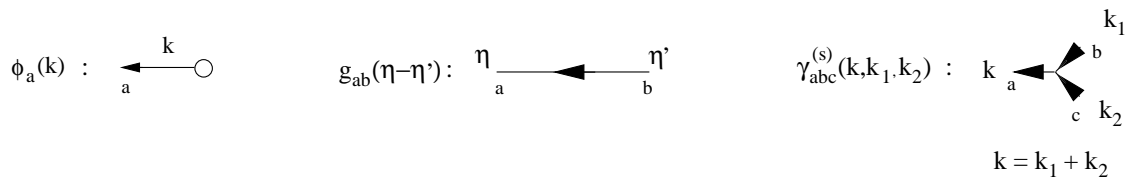}
\caption{Graphical notation of the basic objects in the perturbative expansion, the {\em initial field} $\phi_a$, the {\em linear propagator} $g_{ab}$, and the {\em vertex} $\gamma_{abc}^{(s)}$.}
\label{figure1}
\end{figure}

\begin{figure}[h!]
\begin{center}
\includegraphics[width=0.8\textwidth]{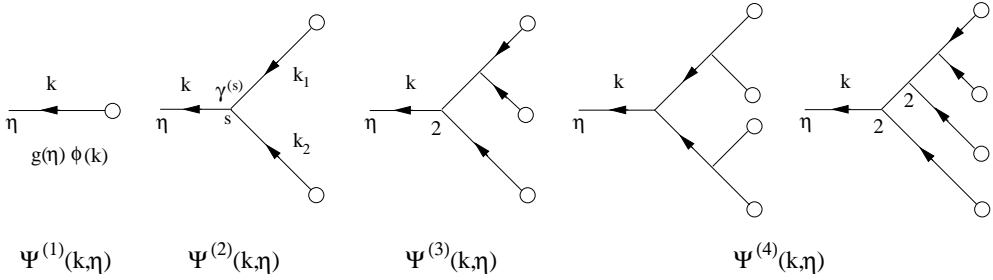}
\caption{Diagrams up to order $n=4$ in the series expansion of $\Psi(\k,\eta)$.}
\label{figure2}
\end{center}
\end{figure}

As discussed [see Eq.~(\ref{eomi})], there are only three basic building blocks, the {\em initial field} $\phi_a$, the {\em linear propagator} $g_{ab}$, and the {\em vertex} $\gamma_{abc}^{(s)}$. Their graphical representation are shown in Fig.~\ref{figure1}. Since we resort to a graphical representation of the basic building blocks rather than derived quantities such as  the kernels (as in done in the diagrammatic approach in standard PT~\cite{Fry84, GGRW86,ScFr96}), the vertices in our diagrams are simpler than in standard PT, they always involve three lines (a consequence of quadratic nonlinearities) instead of a variable number of lines depending on the order in PT. The complication, at first sight, is that the diagrams carry information about time variables when the nonlinear interactions occur, and one integrates over these intermediate times (effectively summing over all possible interactions that happen between the initial conditions and the present time). This information is hidden in standard PT, where time evolution has been already ``integrated out" (standard PT diagrams can be indeed thought as ``collapsing" the diagrams  shown in Fig.~\ref{figure2} below in the time direction). We will see here that this information can be taken advantage of, in fact {\em making possible} the process of resummation.

To obtain the set of diagrams corresponding to the $n$th order term in Eq.~(\ref{seriesexp}) one has to draw all topologically different trees with $n-1$ vertices (branchings) and $n$ initial conditions. Each tree is obtained as follows: from the final time variable $\eta$ we draw a line backwards until it reaches a vertex, where the line bifurcates into two branches, with each of them continuing until they reach another vertex or an initial field at $\eta=0$. If the branching is asymmetric it carries a factor of $2$. This process is repeated at each vertex until all the branches end up in initial fields.

Each diagram with $n-1$ vertices represents an integral contributing to $\Psi^{(n)}$. The rules to obtain these integrals are as follows. Each of the $n$ initial fields $\phi$ is characterized by a momentum $\k_i$. Each branching corresponds to a vertex $\gamma^{(s)}(\k,\k_1,\k_2)$ and represents the interaction of two incoming waves $\k_1$ and $\k_2$ that couple to form one outgoing $\k$ (due to the quadratic nonlinearities in the equations of motion). Each interaction happens at a given time $s_j$ ($0\leq s_j \leq \eta$) and conserves momentum ($\k=\k_1+\k_2$). The branches correspond to linear propagators representing the linear evolution of a given mode $\k_m$ from time variables $s_i$ until $s_j$. Notice that all the arrows throughout the diagram point away from the initial conditions, indicating the direction of forward time evolution. Finally, all the intermediate wave vectors are integrated over as well as all the time variables $s_j$, each between $[0,\eta]$. The diagrams up to $\Psi^{(4)}$ are shown in Fig.~\ref{figure2}. Note that it is {\em the existence of these simple rules for finding the $n$th order term, which are independent of $n$, that will allow us to considerably simplify the resummation of diagrams. }

As an example of these rules let us write explicitly the integrals corresponding to the $\Psi^{(2)}$ and $\Psi^{(3)}$ diagrams in Fig.~\ref{figure2},

\begin{eqnarray}
\Psi^{(2)}_a(\k,\eta)=\int d^3 k_1 \int d^3 k_2 \int_0^{\eta} ds \ g_{ab} (\eta-s) \ \gamma^{(s)}_{bcd}(\k,\k_1,\k_2) \ g_{ce}(s)\phi_e(\k_1) \ g_{df}(s)\phi_f(\k_2),  \\
\Psi^{(3)}_a(\k,\eta)=2 \int d^3 k_1 \int d^3 k_2 \int_0^{\eta} ds \ g_{ab} (\eta-s) \ \gamma^{(s)}_{bcd}(\k,\k_1,\k_2) \ g_{ce}(s)\phi_e(\k_1) \ \Psi^{(2)}_d(\k_2,s).
\end{eqnarray}

The diagrams we construct here are very similar to any field theory with quadratic nonlinearities, perhaps the closest example to ours is that of turbulence where similar methods are well known~\cite{Wyld61,MSR73,LP95}. There is also a recent paper~\cite{Val04} that applies path-integral methods to  gravitational clustering in cosmology.

\section{Statistics}
\label{stats}

\subsection{Initial Conditions}
\label{StatisticsInitialConditions}

As it stands, the integral Equation~(\ref{eomi}) can be thought as an equation for $\Psi_a(\k,\eta)$ in the presence of an ``external source'' or forcing given by the initial conditions $\phi_a(\k)$ (i.e. $\delta_0(\k)$), with prescribed statistics. Here we assume that the initial conditions are Gaussian; the statistical properties of $\phi_a(\k)$ are then completely characterized by its two-point correlator

\beq
\lexp \phi_a(\k) \ \phi_b(\k') \rexp = \delta_{\rm D}(\k+\k') \
u_a u_b\, P_0(k),
\label{2pt}
\eeq
where $P_0(k)$ denotes the initial power spectrum of density fluctuations. According to Wick's theorem, all higher order correlations of an odd number of fields vanish, whereas for an even number there are $(2n-1)!!$ contributions corresponding to all different pairings of the $2n$ fields,

\beq
\lexp \phi_{a_1}(\k_1) \cdots \phi_{a_{2n}}(\k_{2n}) \rexp = \sum_{{\rm all\,pair\,associations}}\  \ \prod_{\rm p\,pairs\,(i,j)} \lexp \phi_{a_i}(\k_i) \ \phi_{a_j}(\k_j) \rexp.
\label{Wick}
\eeq

\begin{figure}[h!]
\begin{center}
\includegraphics[width=0.3\textwidth]{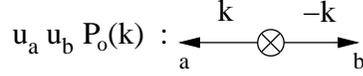}
\caption{Diagrammatic notation for the initial power spectrum.}
\label{figure3}
\end{center}
\end{figure}

When statistics are calculated, pairs of initial fields $\phi_a$ are replaced by the initial power spectrum as one of the basic building blocks. It will be graphically denoted by the symbol shown in Fig.~\ref{figure3}, that arises from gluing a pair of initial fields from Fig.~\ref{figure1}.

\subsection{Power Spectrum}

The goal is to calculate different correlation functions of the final field $\Psi_a(\k,\eta)$, the simplest being its two-point correlator or power spectrum,

\beq
\lexp \Psi_a(\k,\eta) \Psi_b(\k',\eta)  \rexp = \delta_{\rm D}(\k+\k') P_{ab}(\k,\eta).
\label{powerdefinition}
\eeq

Equations~(\ref{2pt}-\ref{Wick}) above give ensemble averages of a product of initial fields $\phi_a$, with the result either vanishing or depending only on the initial power spectrum. This will allow us to develop a perturbative expansion for the final power spectrum in terms of integrals of the initial one, similar to the expansion for $\Psi_a(\k,\eta)$ in terms of $\phi_a(\k)$ described in Eqs.~(\ref{seriesexp}) and (\ref{seriesol}). Replacing the series expansion of Eq.~(\ref{seriesexp}) into Eq.~(\ref{powerdefinition}), we obtain $P_{ab}(\k,\eta)=\sum_{\ell=0}^{\infty} P^{(\ell)}_{ab}(\k,\eta)$ where the $\ell$-loop term is given by (assuming Gaussian initial conditions)
%\beqa
%\lexp \Psi_a(\k,\eta) \Psi_b(\k',\eta)  \rexp &=& \lexp \Psi^{(1)}_a(\k,\eta) \Psi^{(1)}_b(\k',\eta)  \rexp  + \nonumber  \\
%                                        & & \lexp \Psi^{(1)}_a(\k,\eta) \Psi^{(3)}_b(\k',\eta)  \rexp  + \lexp \Psi^{(2)}_a(\k,\eta) \Psi^{(2)}_b(\k',\eta)  \rexp + \lexp \Psi^{(3)}_a(\k,\eta) \Psi^{(1)}_b(\k',\eta)  \rexp + \nonumber   \\
%                                        & & \lexp \Psi^{(1)}_a(\k,\eta) \Psi^{(5)}_b(\k',\eta)  \rexp  + \lexp \Psi^{(2)}_a(\k,\eta) \Psi^{(4)}_b(\k',\eta)  \rexp +  \ldots ,
%\eeqa
%where we used that $\Psi_a^{(n)}$ contains $n$ powers of initial fields and the ensemble average of an odd number of initial fields vanishes. In general, the $n$th order term of the expansion $P_{ab}(\k,\eta)=\sum_{n=1}^{\infty} P^{(n)}_{ab}(\k,\eta)$ satisfies

\beq
 \delta_{\rm D}(\k+\k') P^{(\ell)}_{ab}(\k,\eta) = \sum_{m=1}^{2\ell+1} \lexp \Psi^{(m)}_a(\k,\eta) \Psi^{(2\ell+2-m)}_b(\k',\eta) \rexp.
\eeq

\begin{figure}[h!]
\begin{center}
\includegraphics[width=0.9\textwidth]{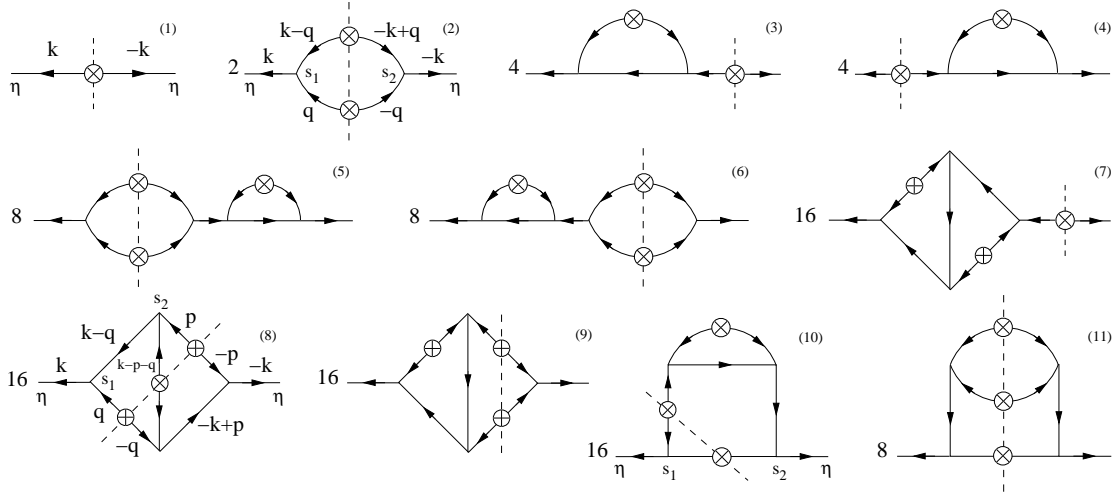}
\caption{Diagrams for the correlation function $P_{ab}(\k,\eta)$ up to two-loops (only 7 out of 29 two-loop diagrams are shown here). The dashed lines represent the points at which the two trees representing perturbative solutions to $\Psi_a$ and $\Psi_b$ have been glued together.}
\label{figure4}
\end{center}
\end{figure}

We can now extend the diagrammatic representation of Section~\ref{DiagrammaticRepresentation} to describe each term in the above expansion. To draw each diagram that contributes to $P_{ab}^{(\ell)}$, we put one of the tree-like diagrams for $\Psi_a^{(m)}$ against one for $\Psi_b^{(2\ell+2-m)}$ with their initial fields facing each other. Next, we pair the initial fields in all possible ways and glue the pairs. Each pair of initial fields with characteristic wave numbers $\k_i$ and $\k_j$ and indices $c$ and $d$ is then converted to an initial power spectrum $\delta_{\rm D}(\k_i+\k_j) \, u_c u_d P_0(\k_i)$. Finally, we do the same with all combinations of one tree of $\Psi_a^{(m)}$ with one of $\Psi_b^{(2\ell+2-m)}$ (i.e. taking into account that after order $n=3$ there is more than one tree at each order). Going through this process one often obtains  the same diagram. Therefore, the final weight of a diagram is given by the numerical factor for each ``independent'' diagram due to its trees, as described in Fig.~(\ref{figure1}), and another factor that takes into consideration the counting of the equal diagrams generated by the pairing process. All diagrams up to one-loop and some of the $29$ two-loop diagrams are shown in Fig.~\ref{figure4} (see \cite{Wyld61} for a full account in the similar case of turbulence).

Two sets of diagrams have not been included in Fig.~\ref{figure4} since they vanish identically. One set is composed by diagrams that contain a sub-diagram linked to the main part by only one linear propagator. It can be shown that the sum of these fragments to all orders gives the ensemble average of the field, $\lexp \Psi_a(\k,\eta)\rexp$, which is zero. For a similar reason the set of unconnected diagrams vanishes.

For future use in the process of resummation, we define here the ``{\em principal cross section}'' of a diagram as the line that splits the diagram into two cutting only through initial power spectra. It represents the points where the two trees have been glued together. It is shown as a dashed line for the different diagrams in Fig.~\ref{figure4}.  Each diagram in Fig.~\ref{figure4} is in one-to-one correspondence with a certain integral. The rules for obtaining these integrals are the same to those described in Section~\ref{DiagrammaticRepresentation}. For example, the diagram $1$ in Fig.~\ref{figure4}, leads to the well known $linear$ power spectrum for growing mode initial conditions ($u_a=(1,1)$)

\beq
P^{(0)}_{ab}(\k,\eta)|_{\rm d 1}=g_{ac}(\eta)\,u_c\,P_0(\k)\,u_d\,g_{bd}(\eta) \longrightarrow P_{\rm L}(\k,a)=a^2(\tau)P_0(\k),
\eeq
and the integral corresponding to diagram $2$ (one loop) in the same figure gives,

\beqa
P^{(1)}_{ab}(\k,\eta)|_{\rm d 2}= 2 \int_0^\eta ds_1 \int_0^\eta ds_2 \int d^3q \, \left[\,g(\eta-s_1)\,\gamma^{(s)}(\k,{\bf q},\k-{\bf q})\,\tilde g(s_1) \tilde g(s_1) \right]_a P_0(q) P_0(|\k-{\bf q}|) \nonumber \\  \left[\,g(\eta-s_2)\,\gamma^{(s)}(-\k,-{\bf q},-\k+{\bf q})\,\tilde g(s_2)\tilde g(s_2) \right]_b. \ \ \ \ \
\eeqa
where we introduced the shorthand notations $\tilde g_a(s)=g_{ab}(s)u_b$ ($={\rm e}^{-s} u_a$ for {\it growing mode} initial conditions) and $[ g\,\gamma^{(s)}\,\tilde g\,\tilde g ]_a = g_{ab}\,\gamma^{(s)}_{bcd}\,\tilde g_c \,\tilde g_d$. As a further example, diagram $8$ (two loops) gives,

\beq
P^{(2)}_{ab}(\k,\eta)|_{\rm d 8}=16  \int d^3q \int d^3p \ {\rm G}_a (\k,\q,{\bf p},\eta) \ {\rm G}_b (-\k,-{\bf p},-\q,\eta) \  P_0(p) \ P_0(q) \ P_0(|\k-{\bf p}-\q|), \nonumber 
\eeq
where we made use of the symmetry with respect to the dashed line in Fig~(\ref{figure4}) diagram 8, and defined,
\beq
{\rm G}_a(\k,\q,{\bf p},\eta)=\int_0^\eta ds_1 \int_0^{s_1} ds_2 \ g_{ab}(\eta-s_1) \ \gamma_{bcd}^{(s)}(\k,\q,\k-\q) \  \tilde g_c(s_1)  g_{df}(s_1-s_2) \ \gamma_{fgh}^{(s)}(\k-\q,{\bf p},\k-{\bf p}-\q) \ \tilde g_g (s_2) \tilde g_h(s_2).  \nonumber
\eeq

\subsection{Non Linear Propagator}
\subsubsection{Definition and Diagrammatic Representation}

Nonlinearities can be thought  as modifying the linear propagator defined in Eq.~(\ref{prop}), leading to {\em propagator renormalization}. The non-linear propagator is defined by

\beq
G_{ab}(k,\eta)\ \delta_{\rm D}(\k-\k') \equiv \left\langle \frac{\delta \Psi_a(\k,\eta)}{\delta \phi_b(\k')}\right\rangle.
\label{DefinitionPropagator}
\eeq
Note that translation invariance requires the condition $\k=\k'$, familiar from two-point statistics. Indeed, translation by some vector $\q$, $\x\rightarrow \x+\q$, gives $G \rightarrow G \exp [i(\k-\k')\cdot \q]$. Strictly speaking, this object is not a propagator in the usual sense of a Green's function, since nonlinearities are involved, i.e. it is {\em not} true that in the nonlinear case $\Psi_a(\k,\eta)=G_{ab}(k,\eta)\phi_b(\k)$. But $G_{ab}$ is the generalization of the linear propagator in the sense that its diagrammatic representation also has one entry and one exit arrow. It also generates all diagrams for the two-point function that can be divided into two subdiagrams by cutting a single initial power spectrum $P_0$, similar to what happens with $g_{ab}$ in the diagram for the linear power spectrum (diag. 1 in Fig. \ref{figure4})

With the help of the Eq.~(\ref{seriesexp}), Eq.~(\ref{DefinitionPropagator}) can be cast as a series expansion

\beq
G_{ab}(k,\eta) = g_{ab}(\eta) + \sum_{n=2}^{\infty} \left\langle \frac{\delta \Psi^{(n)}_a(\k,\eta)}{\delta \phi_b(\k)}\right\rangle,
\label{PropagatorExpansion}
\eeq
where we have explicitly separated the linear part from the non-linear contributions. The fully non-linear propagator represents the ensemble averaged response of the final density and velocity divergence fields to variations in the initial conditions. In more physical terms it quantifies how much information of the initial distribution of a $\k$-mode remains in the final state at the same $\k$, in an ensemble average sense. In paper II~\cite{paper2}, we show that for Gaussian initial conditions $\langle \Psi_a \phi_b \rangle = G_{ac} \langle \phi_c \phi_b \rangle$, therefore $G$ can also be thought as a measure of the cross-correlation between final and initial conditions, or indeed as a genuine propagator (or Green's function) in two-point sense. At very large scales, we expect linear evolution to scale initial conditions by the growth factor; consequently, the non-linear corrections in Eq.~(\ref{PropagatorExpansion}) vanish as $\k\rightarrow 0$. As one approaches the non-linear regime interactions gradually ``erase'' the initial distributions; therefore, we expect $G(\k,\eta)$ to vanish as $\k\rightarrow\infty$. See section~\ref{GZA} below for the behavior of $G$ in the Zel'dovich approximation, the case of the exact dynamics is presented in paper II~\cite{paper2} together with comparisons against numerical simulations.

The series defining the nonlinear propagator can be represented in a diagrammatic fashion. For Gaussian initial conditions, only the odd terms in the expansion for $\Psi_a(\k,\eta)$ in  Eq.~(\ref{PropagatorExpansion}) will contribute to $G_{ab}(\k,\eta)$. The $n$th non-linear correction corresponding to $n$ loops is obtained by a functional derivative of $\Psi^{(2n+1)}$. This can be done diagrammatically in a straightforward way. From each tree characterizing $\Psi^{(2n+1)}(\k,\eta)$ in Fig.~{\ref{figure2} there will be $2n+1$ contributions, each obtained after dropping one initial condition represented by a blank circle (this accounts for the functional derivative). The linear propagator left at that branch will carry a wave number $\k$. Finally the ensemble average of the remaining $2n$ $\phi$'s is done by pairing the fields according to Gaussianity.

Figure~\ref{figure5} shows all diagrams up to two loops for the non-linear propagator. Notice that each diagram has one entering arrow and one exiting, connected by a chain of linear propagators that runs through the diagram without intersecting initial conditions. This path is called the ``{\em principal path}'', and will be used later in the process of resummation.

\begin{figure}[h!]
\begin{center}
\includegraphics[width=0.9\textwidth]{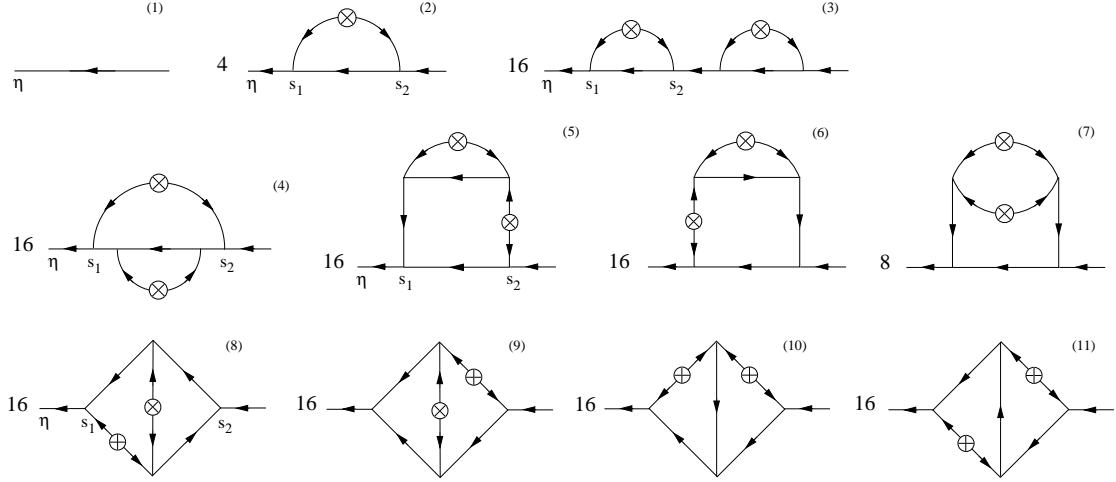}
\caption{Diagrams for the propagator $G(k,\eta)$ up to two loops.}
\label{figure5}
\end{center}
\end{figure}

\subsubsection{A  simple example: Zel'dovich Approximation}
\label{GZA}

To illustrate the behavior of the nonlinear propagator we go back to the discussion in the introduction about the Zel'dovich Approximation. Let us work out the density propagator, $G_\delta \equiv G_{11}+G_{22}$ which assuming growing mode initial conditions follows directly from taking the derivative of the final density with respect to the initial one. For simplicity we consider only the fastest growing mode at each order, so we can use the standard PT kernel representation, Eqs.~(\ref{ptansatz}), (\ref{ec:deltan}) and (\ref{Fn}), and take advantage of the simple form the density field kernel takes in this approximation~\cite{GrWi87}

\beq
F_n (\q_1,...,\q_n)= \frac{1}{n!} \frac{\k\cdot\q_1}{q_1^2}
\ldots \frac{\k\cdot\q_n}{q_n^2}, \label{FZA}
\eeq
where $\k=\q_1+\ldots+\q_n$. Due to Gaussian initial conditions, only odd terms in the series expansion for $\tilde \delta$ contribute, therefore

\beq
G_\delta=a\ \sum_{n=0}^\infty (2n+1)!! \int F_{2n+1}(\k,\q_1,-\q_1,\ldots,\q_n,-\q_n) P_L(q_1,a)d^3q_1 \ldots P_L(q_n,a)d^3q_n,
\eeq
where the term $n=0$ gives just the scale (and growth) factor. Using Eq.~(\ref{FZA}) this series can be summed up right away,

\beq
G_\delta=a \ \sum_{n=0}^\infty \frac{1 }{n!} \Big[-\frac{1}{2} \int \Big(\frac{\k\cdot\q}{q^2}\Big)^2 P_L(q,a)\, d^3q \Big]^n= a \, \exp(-k^2 \sigma_v^2/2) ,
\label{GderivZA}
\eeq
where $\sigma_v^2 \equiv (1/3) \int d^3q P_L(q,a)/q^2$. The extension of this result to the exact dynamics for density and velocity fields and comparison with measurements in numerical simulations is given in~\cite{paper2}.

\subsection{Vertex}

The vertex defined in Eq.~(\ref{vertexdefinition}) is the third basic object of the theory, and is also renormalized due to higher-order nonlinear interactions. We define the symmetric full vertex $\Gamma^{(s)}$ through the relation

\beq
\Big\langle \frac{\delta^2 \psi_a(\k,\eta)}{\delta\phi_e(\k_1)\delta\phi_f(\k_2)} \Big\rangle = 2 \int_0^\eta ds \int_0^s ds_1 \int_0^s ds_2 \ G_{ab}(\eta-s)\, \Gamma^{(s)}_{bcd}(\k,s;\k_1, s_1 ; \k_2, s_2 )\, G_{ce}(s_1)\, G_{df}(s_2) ,
\label{GaRen}
\eeq
thus $\Gamma^{(s)}_{bcd}(\k s , \k_1 s_1 , \k_2 s_2 )=\gamma^{(s)}_{bcd}(\k,\k_1,\k_2)\,\delta_{\rm D}(s-s_1)\,\delta_{\rm D}(s-s_2)+\mbox{``higher-order corrections''}$.

The full vertex depends not only on the two incoming wave vectors but also on the corresponding times when they ``enter'' the interaction. It can be shown that the outgoing wave vector $\k$ is determined by wave vector conservation, $\Gamma^{(s)}_{bcd}(\k,s;\k_1, s_1 ; \k_2, s_2 ) \propto \delta_{\rm D}(\k-\k_1-\k_2)$. The diagrammatic representation up to one loop is shown in Fig.~\ref{figure6}.

\begin{figure}[h!]
\begin{center}
\includegraphics[width=0.8\textwidth]{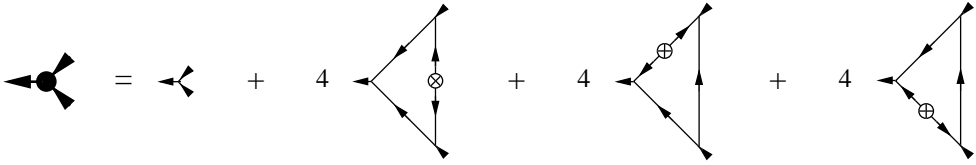}
\caption{Diagrams corresponding to the renormalized vertex $\Gamma$, Eq.~(\protect\ref{GaRen}), up to one loop.}
\label{figure6}
\end{center}
\end{figure}

\section{Renormalized Perturbation Theory (RPT)}
\label{RPT}

\subsection{Non-Linear Propagator}
\label{dyson}

In this section we will perform the first step in the resummation of the series in Eq.~(\ref{PropagatorExpansion}). We divide the diagrams in Fig.~\ref{figure5} into two classes. The {\em Principal Path Reducibles} are the ones that can be split in two disjoint pieces by cutting   one linear propagator belonging to the principal path. Diagram 3 in Fig.~\ref{figure5} is the only example up to two loops. Note that the linking propagator represents the linear evolution of the same Fourier mode that enters and leaves the diagram, and hence it corresponds to the ``independent'' wave vector that is not integrated over. Diagrams that cannot be split in this way are denoted as {\em Principal Path Irreducibles}.

All the {\em Principal Path Irreducible} diagrams start with a linear propagator $g(s_2)$ entering the diagram at $s_2$ and finish with $g(\eta-s_1)$ exiting at $s_1$, with an irreducible structure in between. We define as $\Sigma(\k,s_1,s_2)$ the sum of all these irreducible structures. Thus, we have

\beq
\mbox{Sum of all irreducibles}=g_{ab}(\eta)+\int_0^\eta ds_1 \int_0^{s_1} ds_2 \ \  g_{ac}(\eta-s_1)\,\Sigma_{cd}(\k,s_1,s_2)\,g_{db}(s_2)
\label{step1}
\eeq

The relation $\eta \ge s_1 \ge s_2 \ge 0$ follows from causality along the principal path. All the diagrams for $\Sigma(\k,s_1,s_2)$ up to two loops are shown in Fig.~\ref{figure7}.

\begin{figure}[h!]
\begin{center}
\includegraphics[width=0.6\textwidth]{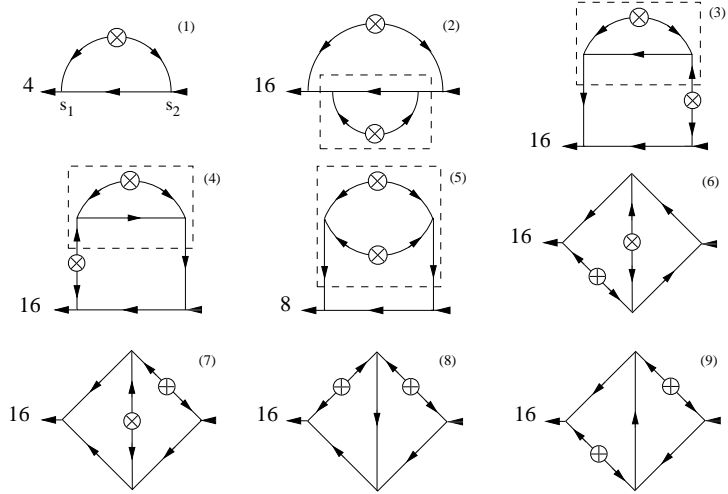}
\caption{Diagrams for $\Sigma(\k,s_1,s_2)$ up to two loops. The dashed lines enclose reducible fragments, see section~\ref{lineres}.}
\label{figure7}
\end{center}
\end{figure}

The {\em Reducible} diagrams are built by two independent fragments linked together by one linear propagator that evolves up to some time, say $s_2$. The diagrams also finish with a linear propagator $g(\eta-s_1)$. This can be clearly seen in diagram 3 in Fig.~\ref{figure5}. There we see that between $s_1$ and $s_2$ there is the first contribution to $\Sigma(\k,s_1,s_2)$ in Fig.~\ref{figure7}. Before $s_2$, there is the first non-linear term in the expansion for $G(\k,s_2)$, that is diagram 2 in the same Figure. The higher-order {\em Reducible} diagrams will have all the higher-order contributions to $\Sigma$ between $s_1$ and $s_2$,  and before $s_2$ all the terms in the series for $G$ except the linear one, $g$. We then write,

\beq
\mbox{Sum of all reducibles}=\int_0^\eta ds_1 \int_0^{s_1} ds_2 \ \  g_{ac}(\eta-s_1)\,\Sigma_{cd}(\k,s_1,s_2)\,[G_{db}(\k,s_2) - g_{db}(s_2)]
\label{step2}
\eeq

Adding up Eqs.~(\ref{step1}) and (\ref{step2}), we arrive at the equation for the non-linear propagator (usually known as Dyson's equation)

\beq
G_{ab}(\k,\eta)=g_{ab}(\eta)+\int_0^\eta ds_1 \int_0^{s_1} ds_2 \ \  g_{ac}(\eta-s_1)\,\Sigma_{cd}(\k,s_1,s_2)\,G_{db}(\k,s_2).
\label{DysonEquation}
\eeq

For further reference we generalize the previous equation to the case where the non-linear propagator evolves from a general time variable $\eta' \neq 0$,

\beq
G_{ab}(\k,\eta,\eta')=g_{ab}(\eta-\eta')+\int_{\eta'}^\eta ds_1 \int_{\eta'}^{s_1} ds_2 \ \ g_{ac}(\eta-s_1)\,\Sigma_{cd}(\k,s_1,s_2)\,G_{db}(\k,s_2,\eta').
\label{GeneralizedDysonEquation}
\eeq

Notice that $G_{ab}(\k,\eta)=G_{ab}(\k,\eta,\eta'=0)$. We will represent the non-linear propagator $G$ by a thick straight line, shown in Fig.~\ref{figure9} together with the graphical representation of Eq.~(\ref{GeneralizedDysonEquation}).

\subsection{Power Spectrum}
\label{wyld}

We now turn to the derivation of the equivalent of Eq.~(\ref{DysonEquation}) for the non-linear power spectrum. Let us first consider in Fig.~\ref{figure4} all the diagrams that contain only one initial power spectrum $P_0(k)$ at the ``principal cross section''. Examples of this type are diagrams $1, 3, 4$ and $7$. They all share the property that at each side of the ``principal cross section'' there is an independent sub-diagram with one entry and one exit arrow connected by a ``principal path''. By summing diagrams of this sort to all orders one ensures to account for all the contributions to the non-linear propagator at both sides of $P_0(k)$ at the ``principal cross section''

\beqa
\mbox{Sum of all diagrams with one initial power}&   &  \nonumber \\
\mbox{spectrum at the ``principal cross section'' }   & = & G_{ac}(\k,\eta) \, u_c \, P_0(k) \, u_d \,  G_{bd}(-\k,\eta).
\label{onepower}
\eeqa
\newline

Consider next diagrams containing two or more initial power spectra at their ``principal cross section''.  From those, diagrams $2, 8, 9, 10, 11$ in Fig.~\ref{figure4} form a special class since they cannot be split in two parts (with more than one $g$ each) by cutting one linear propagator. They are all characterized by an irreducible structure enclosing the ``principal cross section'' finishing with two vertices at, say, $s_1$ to the left and $s_2$ to the right. They continue after each vertex with a linear evolution from $s_i$ to $\eta$ with $g(\eta-s_i)$. We define as $\Phi(\k,s_1,s_2)$ the sum of all these irreducible structures. All diagrams for $\Phi$ up to two loops are shown in Fig.~\ref{figure8}.

The remaining diagrams also contain a contribution to $\Phi(\k,s_1,s_2)$ around the ``principal cross section'', but either to the left of $s_1$ or to the right of $s_2$ they have a non-linear contribution to the non-linear propagator $G(\k,\eta,s_i)$ (diagrams $5$ and $6$ in Fig.~\ref{figure4} respectively). Contributions to the non-linear propagator at both sides of $\Phi$ appear for the first time in three-loop diagrams.

After adding the diagrams described above to all orders, we arrive at

\beqa
\mbox{Sum of all diagrams with two or more initial power} &&       \label{twopowers}   \\
\mbox{spectra at the ``principal cross section'' \ \ \ }  = &&  \int_0^\eta  ds_1 \int_0^\eta ds_2 \, G_{ac}(\k,\eta,s_1) \, \Phi_{cd}(\k,s_1,s_2) \, G_{bd}(-\k,\eta,s_2).  \nonumber
%\label{twopowers}
\eeqa

Adding Eqs.~(\ref{onepower}-\ref{twopowers}) we arrive at the integral equation for the non-linear power spectrum,

\beq
P_{ab}(k,\eta)=G_{ac}(\k,\eta) \, u_c \, P_0(k) \, u_d \,  G_{bd}(-\k,\eta)+\int_0^\eta  ds_1 \int_0^\eta ds_2 \, G_{ac}(\k,\eta,s_1) \, \Phi_{cd}(\k,s_1,s_2) \, G_{bd}(-\k,\eta,s_2). 
\label{WyldEquation}
\eeq

For further reference we extend the previous analysis to the case of a two-point correlation of final fields at different times

\beq
\lexp \psi_a(\k,\eta) \ \psi_b(\k',\eta') \rexp = \delta_{\rm D}(\k+\k') \ P_{ab}(k,\eta,\eta').
\label{2ptdifftimes}
\eeq

The equivalent of Eq.~(\ref{WyldEquation}) for $P_{ab}(k,\eta,\eta')$ reads

\beq
P_{ab}(k,\eta,\eta')=G_{ac}(\k,\eta) \, u_c \, P_0(k) \, u_d \,  G_{bd}(-\k,\eta')+\int_0^\eta  ds_1 \int_0^{\eta'} ds_2 \, G_{ac}(\k,\eta,s_1) \, \Phi_{cd}(\k,s_1,s_2) \, G_{bd}(-\k,\eta',s_2) .
\label{GeneralizedWyldEquation}
\eeq

\begin{figure}[h!]
\begin{center}
\includegraphics[width=0.6\textwidth]{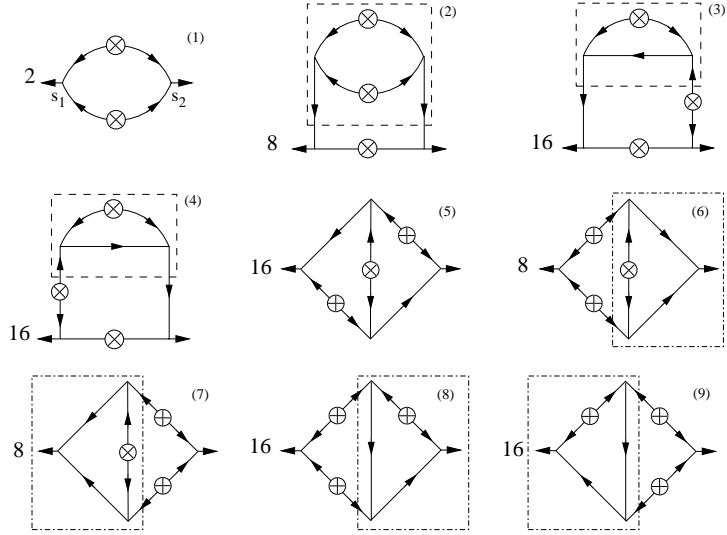}
\caption{All diagrams for $\Phi$ up to two loops. The dashed lines enclose reducible fragments for propagator and power spectrum renormalization. The dash-dotted lines enclose reducible fragments that lead to vertex renormalization, see section~\ref{lineres} for a detailed discussion.}
\label{figure8}
\end{center}
\end{figure}

This equation is shown diagrammatically in Fig.~\ref{figure9}, where we represent the non-linear power spectrum by a thick line with a filled circle at the middle.

\begin{figure}[h!]
\begin{center}
\includegraphics[width=0.6\textwidth]{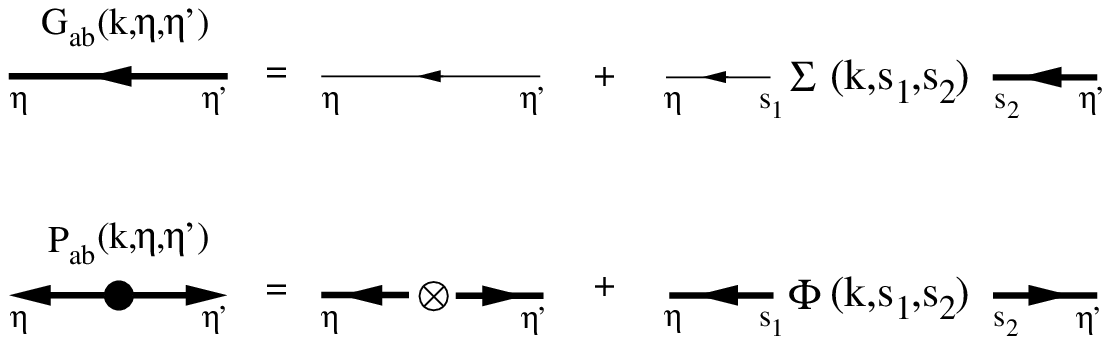}
\caption{Integral Equations~(\ref{GeneralizedDysonEquation}) and (\ref{GeneralizedWyldEquation}) for the non-linear propagator and the non-linear power spectrum.}
\label{figure9}
\end{center}
\end{figure}

It is beyond of the scope of this paper to consider higher-order correlations in detail, but Fig.~\ref{figure12} shows the one-loop diagrams (after renormalization) for the bispectrum
\beq
\lexp \Psi_a(\k_1,\eta) \Psi_b(\k_2,\eta) \Psi_c(\k_3,\eta) \rexp = \delta_{\rm D}(\k_1+\k_2+\k_3)\ B_{abc}(\k_1,\k_2,\k_3,\eta),
\label{bispectrumdefinition}
\eeq
\begin{figure}[h!]
\begin{center}
\includegraphics[width=0.75\textwidth]{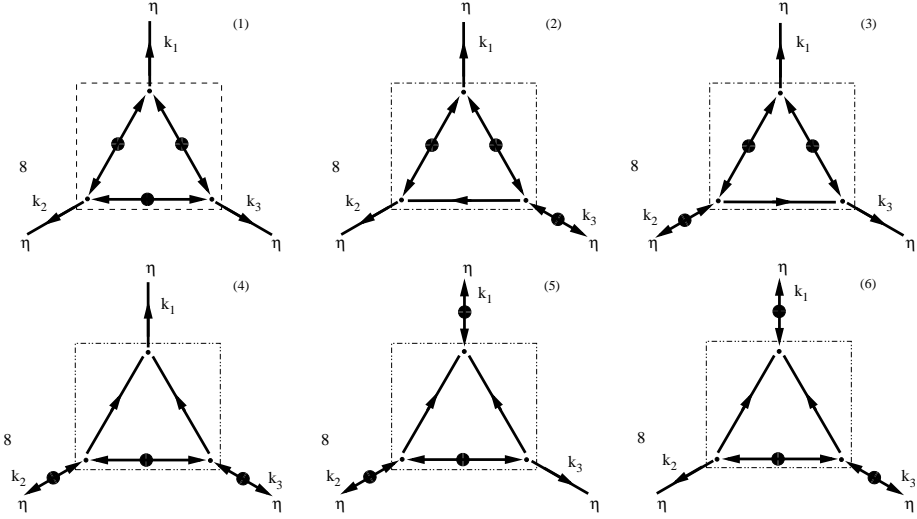}
\caption{Resummed diagrams for the bispectrum $B_{abc}(\k_1,\k_2,\k_3,\eta)$ up to one loop. All the remaining contributions are obtained by cyclic permutations of ($\k_1,\k_2,\k_3$) on diagrams $2$ through $6$.}
\label{figure12}
\end{center}
\end{figure}

\subsection{Further Resummations}
\label{lineres}

In Sections \ref{dyson} and \ref{wyld}, we derived equations that allow the renormalization of the propagator and power spectrum given by the kernels $\Sigma$ and $\Phi$. However the kernels themselves were functions of the linear propagator and linear power spectrum. In this section we complete the resummation procedure by renormalizing $\Sigma$ and $\Phi$, which means that we find expansions for these quantities in terms of non-linear propagators and non-linear power spectra. We will also carry out a vertex renormalization for $\Phi$. 

We begin by noticing that all diagrams for both $\Sigma$ and $\Phi$, in Figs.~\ref{figure7} and \ref{figure8} respectively, contain two ending nodes. For $\Sigma$ there is one entry point and one exit, while for $\Phi$ both are exits. In those figures we define as a ``reducible fragment'' any part of a diagram that can become disjoint from the main part by cutting two linear propagators. We refer as ``main part" of a diagram to the one containing the two ending points and consider the two linking propagators as belonging to the reducible fragment. Reducibles fragments are shown, enclosed by dashed lines, in diagrams $2, 3, 4, 5$ for $\Sigma$ in Fig.~\ref{figure7}, and in diagrams $2, 3, 4$ for $\Phi$ in Fig.~\ref{figure8}.

Notice that there are two types of reducibles fragments, which we call type G and type P, since they lead to propagator renormalization (G) and power spectrum renormalization (P). Type-G fragments contain a principal path, with one linear propagator entering the fragment and another exiting it. Examples are the fragments in diagrams $2, 3, 4$ for $\Sigma$ and $3, 4$ for $\Phi$. Type-P fragments contain a local principal cross section, with two linear propagators going out of the fragment, like the one in diagrams $5$ for $\Sigma$ and $2$ for $\Phi$.

Let us analyze first diagrams with type-G fragments.
In the series for $\Sigma$, we see that diagram $2$ can be generated by replacing $g(s_1-s_2)$ in diagram $1$ by the first non-linear contribution to $G(\k,s_1,s_2)$ (shown in diagram $2$ in Fig.~\ref{figure5}). Likewise diagrams $3, 4$ can be obtained out of diagram $1$ by replacing linear by non-linear propagators. In the series for $\Phi$, diagrams $3$ and $4$ can be obtained by substituting one linear propagator in diagram $1$ by the first non-linear correction to $G(\k,s)$.

Next, we turn to diagrams containing type P fragments.
In Fig.~\ref{figure7}, diagram $5$ can be obtained by replacing the linear power spectrum in diagram $1$ by the first non-linear contribution to $P(\k,s_1,s_2)$ (shown in diagram $2$ of Fig.~\ref{figure4}). Similarly, in Fig.~\ref{figure8} diagram $2$ is equivalent to diagram $1$ with the linear propagator replaced by the one-loop correction.

It is possible to show that the sum of type-G fragments is the complete non-linear correction to the non-linear propagator, while the sum of type-P fragments equals the complete non-linear correction to the non-linear power spectrum. This completes the resummation procedure for $\Sigma$ in terms of non-linear propagators and power spectra. The resummed diagrams, up to two loops, are shown in Fig.~\ref{figure10}.

The renormalization of $\Phi$ can be taken one step further by performing the resummation of the vertex. Diagrams $6$ and $7$ in Fig.~\ref{figure8} can be obtained from diagram $1$ by replacing the corresponding tree-level vertex $\gamma$ by the first one-loop contribution to the full vertex, $\Gamma$, shown in Fig.~\ref{figure6}. Similarly, diagrams $8$ and $9$ are reproduced with the help of the second and third one-loop diagrams in Fig.~\ref{figure6}. 

With the renormalization procedure described in this paper for power spectrum, propagator and vertex we were able to account for all the diagrams in the series for $\Phi$ up to two loops, with the renormalized series given in Fig.\ref{figure11}. In \cite{Wyld61}, the author carried out the same procedure one order higher, and could describe all the series for $\Phi$ up to three loops, with $19$ new three-loop irreducible diagrams added to those in Fig. \ref{figure11}. However, in \cite{MSR73,LP95} it has been pointed out the necessity to introduce three different vertices to account for all diagrams, in the series of both $\Sigma$ and $\Phi$, after two loops. A detailed discussion of vertex renormalization is beyond the scope of this paper, and will be discussed elsewhere. 
In our companion paper, \cite{paper2}, we get around this issue for the case of the non linear propagator (i.e. $\Sigma$) by summing the bare perturbation series in Fig.~\ref{figure5}.

In summary, it is possible  to obtain the series for $\Phi$, at least up to two loops, with only the corresponding irreducibles diagrams written in terms of non-linear propagators, power spectra, and full vertices. The first diagram in Fig.~\ref{figure8} is the irreducible at the one-loop level, while the fifth is the only irreducible at the two-loop level. Figure \ref{figure11} shows the final ``renormalized'' results for $\Phi$.

It is worth mentioning that the same resummation procedure can be done for the full vertex in Fig.~\ref{figure6}. The ``renormalized'' series for $\Gamma^{(s)}$ is, to one loop, still given by the diagrams in Fig.~\ref{figure6}, but with all quantities changed by their fully non-linear counterparts \cite{Wyld61}.

\begin{figure}[h!]
\begin{center}
\includegraphics[width=0.6\textwidth]{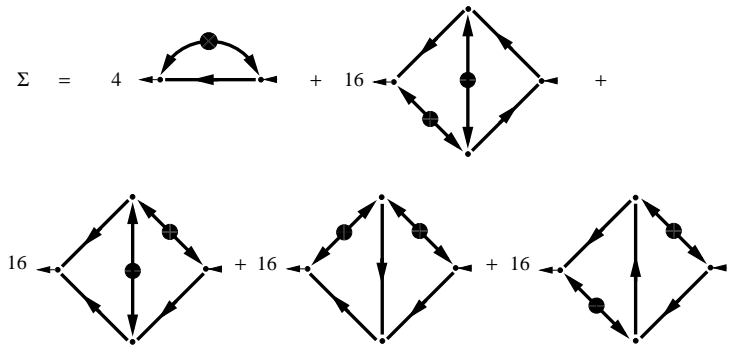}
\caption{Renormalized series expansion for $\Sigma$ up to two loops.}
\label{figure10}
\end{center}
\end{figure}

\begin{figure}[h!]
\begin{center}
\includegraphics[width=0.6\textwidth]{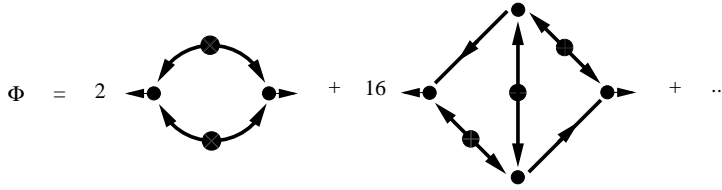}
\caption{Renormalized series expansion for $\Phi$ up to two loops.}
\label{figure11}
\end{center}
\end{figure}

\section{RPT and the halo model}
\label{halos}

The expression in Eq.~(\ref{WyldEquation}) corresponding to the bottom diagram in Fig.~\ref{figure9} has some interesting similarities with the expression for the power spectrum in the halo model~\cite{HaloModel}. In RPT, the power spectrum is written as the sum of two terms, one that dominates at large scales (similar to the 2-halo term) and the other (involving $\Phi$) that dominates at small scales (similar to the 1-halo term). The RPT expression for the Zel'dovich approximation, shown in the right panel in Fig.~\ref{PTvsRPT} and Eq.~(\ref{PrptKernels}) makes the analogy very clear: the $n=1$ term corresponds to the linear power spectrum modulated by a scale-dependent ``bias factor" due to the renormalized propagator that becomes suppressed in the nonlinear regime; whereas the $n\geq2$ sum (weighted by the square of the PT kernels) resembles the sum over halo masses (weighted by the square of the halo profiles) in the 1-halo power spectrum.

The analogy can be taken a bit further by calculating the nonlinear propagator in the halo model, where the density can be written as a sum over halos of mass $m_i$ and profile $u_{m_i}(\x)$ (normalized so that $\int d^3x\, u_m(\x)=1$)

\beq
\rho(\x)= \sum_i m_i\, u_{m_i}(\x-\x_i)=\int m\, dm\, d^3 x'\, u_m(\x-\x') \sum_i \, \delta_{\rm D}(m-m_i) \, \delta_{\rm D}(\x_i-\x'),
\eeq
where the mass function is obtained from the average $\langle \sum_i \delta_{\rm D}(m-m_i) \delta_{\rm D}(\x_i-\x') \rangle = n(m)$ and the mean density is $\bar{\rho} = \int m\, dm\, n(m)$. As we show in~\cite{paper2} the nonlinear propagator can also be calculated for Gaussian initial conditions from cross-correlating final and initial fields, and for growing mode initial conditions the density propagator becomes $G_\delta(k)=\langle \delta(\k) \delta_0(\k')\rangle /\langle \delta_0(\k) \delta_0(\k')\rangle $, which reduces to the growth factor $D_+$ in linear theory ($\delta_L=D_+ \delta_0$). The correlation between halo positions and the initial density field is not known in detail, but following the standard assumptions in the calculation of the 2-halo term, we can model it introducing the halo linear bias factor $b_1(m)$

\beq
 \sum_i \, \delta_{\rm D}(m-m_i)\, \delta_{\rm D}(\x_i-\x')\simeq n(m)\,  \Big[1+b_1(m) \, \delta_L(\x')\Big],
 \label{assump}
 \eeq
which implies, after straightforward algebra, that in the halo model,

\beq
G_\delta(k)=  \frac{D_+}{\bar{\rho}} \int m\, dm\, n(m)\, b_1(m)\ u_m(k) = D_+\, b_1(k),
\label{Ghalo}
\eeq
in other words, propagator renormalization in RPT exactly corresponds to the standard (scale-dependent) bias in the framework of the halo model, making the analogy between the 2-halo term and the first term in  Eq.~(\ref{WyldEquation})  exact. In practice the suppression from the propagator in RPT is stronger than that in the halo model [due to the halo profile in Eq.~(\ref{Ghalo})], however this is a simplification of Eq.~(\ref{assump}) where nonlinear halo bias and exclusion effects are ignored.

Regarding the 1-halo term, it is difficult to connect it formally to the mode-coupling kernel $\Phi$ in Eq.~(\ref{WyldEquation}) because there is no perturbative description of such a term and $\Phi$ does not appear to have a simple expression apart from obeying Eq.~(\ref{WyldEquation}). However, physically, the relationship is obvious: the second term in Eq.~(\ref{WyldEquation}) describes mode-mode coupling and so does the 1-halo term (given the usual assumptions where the 2-halo term is proportional to the linear spectrum). In this sense the correspondence between the description of the power spectrum in RPT and the halo model is exact, although the ingredients are very different.

\section{Conclusions}
\label{conclude}

We developed a new way of looking at cosmological perturbation theory, which makes possible a well-controlled description of gravitational clustering at non linear scales.  This formalism follows the growth of structure as it develops in time, decomposed into linear propagation plus interactions, summing over all possibilities. Although at first sight such time-decomposition seems more complicated than in standard perturbation theory (where time evolution is already integrated out up to the present time), we showed that one can take advantage of this extra information by finding (based on topological considerations)  infinite subset of diagrams that can be resummed and identified with physical quantities.

The most important aspect of this resummed theory, which we call renormalized perturbation theory (RPT), is that the nonlinear (renormalized) propagator (which enters as a key ingredient in the calculation of correlation functions) is a strong function of scale, decaying nearly exponentially at nonlinear scales, see~\cite{paper2} for measurements in numerical simulations. When the perturbative expansion of the nonlinear power spectrum is divided into terms that sum up to the renormalized propagator plus terms that describe mode-mode coupling, the resulting (renormalized) perturbation theory is well behaved in the sense that each term dominates in a narrow range of scales and is suppressed otherwise. We also showed that this description turns out to be analogous to that in the halo model.

Future work is needed to decide whether this approach will be fruitful in quantitative terms; for example, by comparing predictions for the nonlinear power spectrum against N-body simulations. In a companion paper~\cite{paper2} we present a detailed analysis of the resummation of the propagator and comparison of it against numerical simulations. The calculation of the nonlinear power spectrum in RPT requires the inclusion of many terms (loops) in the mode-coupling series to cover an interesting range of scales, which at present appears as a nontrivial task. However, there are symmetries (e.g. Galilean invariance, see~\cite{ScFr96}) that connect the resummation of the mode-coupling series with that of the propagator, which one might be able to take advantage of. This issue deserves further work and will be discussed elsewhere~\cite{paper3}.

Finally, one might wonder whether the single-stream approximation made so far in our formalism will break down before one can get any interesting results into the nonlinear regime~\footnote{Extension of our approach to the Vlasov equation is certainly possible, see~\cite{Val04} for application of path-integral methods to the Vlasov-Poisson system.}. Although this is a possibility, an educated guess is that such effects do not show up until scales well in the nonlinear regime, if one uses as a proxy for when multistreaming becomes important the scale at which the power in the vorticity of the velocity field equals that in the divergence~\cite{Sco00}. This happens (not surprisingly, since vorticity should slow down the growth of power) at roughly the same scale of the ``virial turnover" of the nonlinear power spectrum, i.e. $k \simeq 1 \kvecMpc$ for CDM models at $z=0$. Having a robust prediction of the nonlinear power spectrum as a function of cosmological parameters up to such scales would be very useful for applications in many aspects of large-scale structure, e.g. weak gravitational lensing surveys.

\acknowledgments

We thank G. Gabadadze, A. Gruzinov, D.W. Hogg, S. Pueblas, and E. Sefusatti,  for useful discussions.

\end{document}